\documentclass[12pt]{iopart}
\usepackage{graphicx}

\begin{document}

\title{Mass distributions for nuclear disintegration from fission to evaporation}

\author{N.~Eren$^1$, N.~Buyukcizmeci$^1$ and R.~Ogul$^{1,2}$}

\address{$^1$Department of Physics, University of Sel\c{c}uk,
TR-42079 Konya, Turkey}
\address{$^2$Gesellschaft f{\"u}r Schwerionenforschung
mbH, D-64291 Darmstadt, Germany}

\ead{eren@selcuk.edu.tr}
\begin{abstract}
By a proper choice of the excitation energy per nucleon we analyze
the mass distributions of the nuclear fragmentation at various
excitation energies. Starting from low energies (between $0.1$ and
$1$ MeV/nucleon) up to higher energies about $12$ MeV/n, we
classified the mass yield characteristics for heavy nuclei
($A>200$) on the basis of Statistical Multifragmentation Model.
The evaluation of fragment distribution with the excitation energy
show that the present results exhibit the same trend as the
experimental ones.
\end{abstract}



\vspace{0.5cm}
\section{Introduction}

Properties of nuclear reactions have been under investigation for
several decays. Experimental and theoretical studies on nuclear
reactions are very important not only for the context of nuclear
physics but also for our understanding of astrophysical events,
such as supernova explosion and formation of neutron stars
\cite{Botvina04}. At low excitation energies up to $1$ MeV per
nucleon, the fission of compound nucleus or its evaporation to
small particles have been observed both theoretically and
experimentally. At the excitation energies in between $1$ and $3$
MeV per nucleon one may observe a U-shape distribution
corresponding to partitions with a few small fragments and one big
residual fragment, which looks like an evaporative emission. At
excitation energies greater than $3$ MeV per nucleon one may
observe a mass distribution of intermediate mass fragments
(nuclear multifragmentation). During this process, the hot and
compressed nuclei tend to expand in thermodynamical equilibrium.
Then they enter the region of subsaturation densities, where they
become unstable to density fluctuations and break up into the
fragments. In this case, it is believed that a liquid-gas phase
transition is manifested \cite{SMM,ogul,nihal,Bot3}. At higher
energies $8-9$ MeV/n, big fragments disappear, and an exponential
fall of the mass distribution with mass number A may be seen.

By a proper choice of the excitation energy per nucleon we analyze
the mass distributions of the nuclear fission of heavy elements
such as  $^{238}$U and $^{226}$Ra at various excitation energies
on the basis of Statistical Multifragmentation Model (SMM). We
have also discussed the mass distribution of the nuclear
fragmentation at higher energies up to $12$ MeV/n.

\vspace{0.5cm}
\section{Calculations and results}

Nuclear fission is one of the most interesting channel of
de-excitation of heavy nuclei at low excitation energies up to
$~1$ MeV/nucleon, at which nuclear density exhibits small
fluctuations around the equilibrium density $0.16$ fm$^{-3}$. In
order to study compound nucleus fission one should consider a
correct physical description in the light of liquid drop model
with shell effects at various deformation modes, where the
determination of angular momentum and dynamical variables becomes
very important. In our calculations, we consider the Bohr-Wheeler
approach \cite{Bohr} in SMM, where the partial width of the
compound nucleus is given by
\begin{equation}
    \Gamma_{f}=1 / [{2 \pi \rho_0(E^*)}] \int_{0}^{E^*-B_f} \rho_f
    ({E^*-B_f-E}) dE
\end{equation}
where $ \rho_0$ is the level density of compound nucleus, $
\rho_f$ the level density at the saddle point, $ E^*$ the
excitation energy for the nucleus, and $B_f$ the height of the
fission barrier. Here, the shell effect on the level densities of
the nucleus can be neglected at high energies \cite{Cherepanov}.
The fission barrier is determined by Myers and Swiatecki
\cite{Myers}, and we use the results in Ref. \cite{Iljinov} for
the level density at saddle point.

\begin{figure}[htbp]
\begin{center}
\includegraphics[width=7cm,height=8cm]{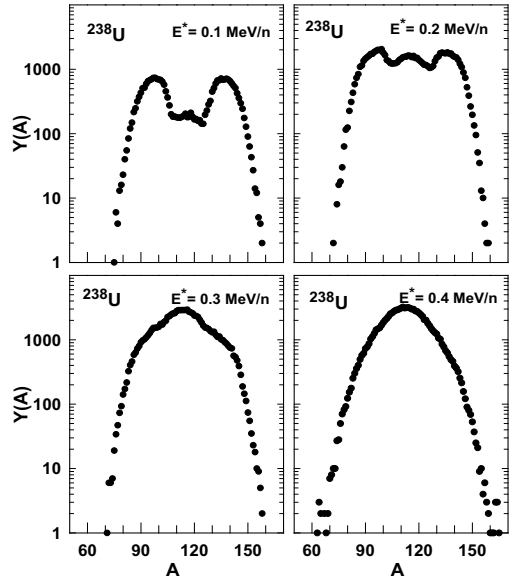}
\end{center}
\caption{\small{Mass yield distributions for the fission of
$^{238}$U at various excitation energies between $0.1$ and $0.4$
MeV per nucleon within SMM.}}
\end{figure}
For orientation, we have shown the results of our calculations at
low excitation energies for the fission of $^{238}$U in Fig. 1. In
this figure, one may see a double peaked distribution at an
excitation energy of $0.1$ MeV/n on the upper-left panel, which
can be interpreted as asymmetric fission, where the mass ratio of
the most probable fragments is about $3$ to $2$. In the
upper-right panel one may see a triple peaked mass distribution,
which is usually interpreted in terms of a single symmetric peak
and a double-peaked distributions. In the lower panels we show the
characteristic of symmetric fission, associated with higher
excitation energies. The evaluation of mass yield distribution
with the excitation energy show that the present results are in
agreement with those obtained experimentally in the reactions
induced by protons and deuterons on $^{226}$Ra \cite{Perry}.

\begin{figure}[htbp]
\begin{center}
\includegraphics[width=8.5cm,height=6cm]{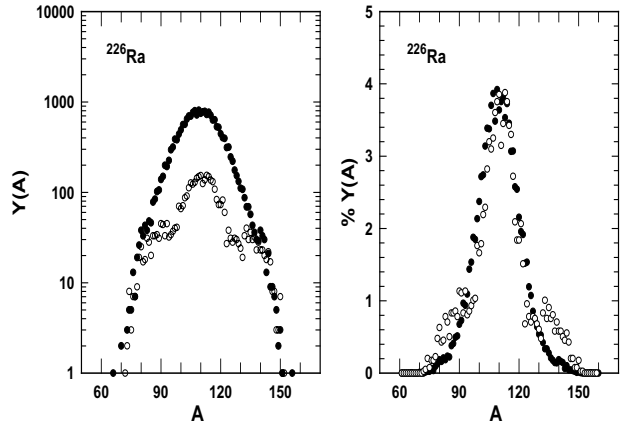}
\end{center}
\caption{\small{The left panel shows the mass yields of the
fission of $^{226}$Ra as a triple peaked mass distribution at
$0.1$ MeV/n (denoted by the open circles), and a single symmetric
distribution at $0.2$ MeV/n (the full circles). The right panel
illustrates the percentage of the same distributions.}}
\end{figure}
It is also observed that the proportion of the symmetric component
increases with increasing excitation energy, in agreement with
experimental results. In order to make a comparison with
\cite{Perry}, we have chosen the nucleus $^{226}$Ra, and showed
the results of our calculations in Fig. 2, from which one may see
a triple peaked mass distribution at $0.1$ MeV/n, and a single
symmetric distribution at $0.2$ MeV/n. The triple peaked
distribution can be interpreted in terms of a symmetric fission
associated with higher excitation energies, and an asymmetric
component associated with lower excitation energies. In the right
top and bottom panels, we also show the results in percentage of
the variation of the mass yield with mass number A. These results
are very similar to the obtained experimentally in Ref.
\cite{Perry}. We also review the fragmentation phenomena at higher
excitation energies on the basis of SMM. According to SMM, one
assumes a micro-canonical ensemble of breakup channels, and the
system should obey the laws of conservation of energy E*, mass
number A and charge number Z. The probability of generating any
breakup channels is assumed to be proportional to its statistical
weight as
\begin{equation}
    W_j \propto exp(S_j(E^*,A,Z))
\end{equation}
where $S_j$ denotes the entropy of a multifragment state of the
breakup channel j. The breakup channels are generated by Monte
Carlo method according to their weights. Light fragments with mass
number   and charge number   are considered as stable particles
(nuclear gas) with masses and spins taken from the nuclear tables.
Only translational degrees of freedom of these particles
contribute to the entropy of the system. Fragments are treated as
heated nuclear liquid drops, and their individual free energies
are parameterized as a sum of the bulk, surface, Coulomb and
symmetry energy contributions
\begin{equation}
F_{A,Z}=F_{A,Z}^B+F_{A,Z}^S+E_{A,Z}^C+E_{A,Z}^{sym}.
\end{equation}
The bulk contribution is given by $F_{A,Z}^B=(W_0-T^2/
\varepsilon_0)A$, where T is the temperature, the parameter $
\varepsilon_0$ is related to the level density, and $W_0=16$ MeV
is the binding energy of infinite nuclear matter. Contribution of
the surface energy is $F_{A,Z}^S=B_0 A^{2/3}[(T_{\rm
c}^2-T^2)/(T_{\rm c}^2+T^2)]^{5/4}$, where $B_0=$ 18 MeV is the
surface coefficient, and $T_{\rm c}=18$ MeV the critical
temperature of the infinite nuclear matter. Coulomb energy
contribution is $E_{A,Z}^C=cZ^2/A^{1/3}$, where c denotes the
Coulomb parameter obtained in the Wigner-Seitz approximation,
$c=(3/5)(e^2/r_0)(1-( \rho / \rho_0)^{1/3})$, with the charge unit
e, $r_0=1.17$ fm, and $\rho_0$ is the normal nuclear matter
density (0.15 fm$^{-3}$). And finally, the symmetry term is
$E_{A,Z}^{sym}= \gamma (A-2Z)^2/A$, where $\gamma =25$ MeV is the
symmetry energy parameter. All the parameters given above are
taken from the Bethe-Weizsäcker formula and correspond to the
assumption of isolated fragments with normal density in the
freeze-out configuration.
\begin{figure}[htbp]
\begin{center}
\includegraphics[width=7cm,height=6cm]{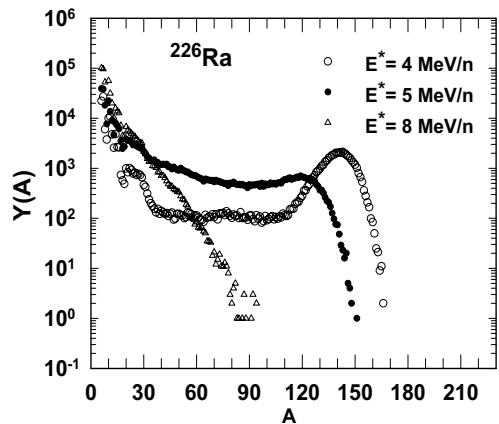}
\end{center}
\caption{\small{Mass yield distributions for the fragmentation of
$^{226}$Ra nucleus, at excitation energies of E* = $4, 5$ and $8$
MeV/n, within SMM.}}
\end{figure}

In Fig. 3 we show the results of our calculation for typical mass
distributions for the fragmentation of $^{226}$Ra nucleus, at
excitation energies of E* = $4$, $5$ and $8$ MeV/n. One may see
from this figure that at  an excitation energy of E* = $4$ MeV/n
(corresponding temperature T $\leq 5$ MeV), there is a U-shape
distribution corresponding to partitions with few small fragments
and one big residual fragment. In the so called transition region
(T $\approx 5-6$ MeV), however, variation of T with E* exhibits a
plateau-like behavior that can be interpreted as a sign of
liquid-gas phase transition in the system
\cite{Agostino,Trautmann,Pochodzalla}, and one observes a smooth
transformation for E* = $5$ MeV/n in the same figure. At high
temperatures (T $\geq 6$ MeV) for E* = $8$ MeV/n, the big
fragments disappear and an exponential-like fall-off is observed.
For higher excitation energies (E* $ > 12$ MeV/n) the nuclei tend
to evaporate to nucleons and small fragments. All these results
are in good agreement with experimental data
\cite{Bot3,Elliott,Schmelzer,Reuter,Mahi}. In view of these
theoretical results it is instructive to demonstrate the
possibility of the application of such approaches for the analysis
of experimental data for nuclear reactions and astrophysical
studies \cite{Botvina04,mish}.

\vspace{0.5cm}
Many helpful discussions with A.S. Botvina are
gratefully acknowledged. The authors thank the Board of Scientific
Research Project of Sel\c{c}uk University(BAP). R.O. thanks GSI
for hospitality, where part of this work was carried out.

\end{document}